\documentstyle[aps,multicol,psfig]{revtex}

\begin{document}

\newcommand{\be}{\begin{eqnarray}}
\newcommand{\ee}{\end{eqnarray}}
\newcommand{\nn}{\nonumber}

\draft

\title{Interface dynamics in Hele-Shaw flows with centrifugal forces.\\
 Preventing cusp singularities with rotation}

\author{F.X. Magdaleno, A. Rocco and J. Casademunt}
\address{Departament d'Estructura i Constituents de la Mat\`{e}ria, Facultat de F\'{\i}sica, \\
Universitat de Barcelona, Av. Diagonal 647, E-08028 Barcelona, Spain}

\date{\today}

\maketitle

\begin{abstract}
A class of exact solutions of Hele-Shaw flows without surface tension 
in a rotating cell is reported.
We show that the interplay between injection and rotation modifies 
drastically the scenario of formation of finite-time 
cusp singularities. For a subclass of solutions, we show that, 
for any given initial condition, there exists a critical rotation rate 
above which cusp formation is prevented. 
We also find an exact sufficient condition 
to avoid cusps simultaneously for all initial conditions. This condition 
admits 
a simple interpretation related to the linear stability problem. 
\end{abstract}


\begin{multicols}{2}

The dynamics of the interface between viscous fluids confined in a 
Hele-Shaw cell\cite{Saffman58,Bensimon86,Segur91} has received
attention for several decades from physicists, mathematicians and 
engineers. In particular it has played a central role in the context 
of interfacial pattern formation\cite{Pelce88,Kessler88}.
As a free 
boundary problem it has the particular interest that explicit time-dependent 
solutions can often be found in the case with no surface 
tension\cite{Sarkar84,Howison85,Mineev94,Dawson98}. 
As a consequence, the issue of the role of surface tension as a singular 
perturbation in the interface
\it dynamics \rm has 
received increasing attention  
\cite{Dai91,Tanveer93,Siegel96,Magdaleno98,Feigenbaum99}
for its potential relevance to a broad class of problems. 
However, to what extent 
the physics of the real problem (with finite surface tension) 
is captured, even
at a qualitative level, by the known solutions is still poorly understood.

As an initial-value problem, the zero surface tension case is 
known to be ill-posed\cite{Tanveer93}. 
An important aspect related to this fact is that 
some smooth initial conditions develop finite-time singularities in the 
form of cusps of the interface\cite{Sarkar84,Howison85}. 
After this blow-up of the solution, the 
time evolution is no longer defined. Generation of finite-time singularities
is in itself interesting in connection with other singular perturbation 
problems in fluid dynamics, such as in the case of the Euler equations.
In the present problem, surface tension acts as the 
natural regulator curing this singular behavior, but unfortunately the problem 
with surface tension is much more difficult and 
usually defies the analytical treatment. Motivated by 
this fact, and inspired by recent experiments on rotating Hele-Shaw
cells\cite{Carrillo96,Carrillo99}, 
we address here the 'perturbation' of the original free boundary 
problem by the presence of a centrifugal field. This new 
ingredient enriches the problem 
in a nontrivial way but, as we will see, it
still admits explicit solutions, and may thus 
lead to new analytical insights, both in understanding the generation
of finite-time singularities, 
and in elucidating 
the dynamical mechanisms of Laplacian growth with and without surface tension.
In fact, although the new dimensionless parameter introduced by 
rotation does not fully regularize the problem, we show that it
may prevent the occurrence of finite-time singularities, allowing for 
smooth, nonsingular solutions for the entire time evolution. This fact
alone enlarges the class of (nontrivial) 
exact solutions without surface tension
which are potentially relevant to the physically realizable situations.

We study an interface between a fluid with viscosity $\mu$ and density $\rho$ 
and one with zero viscosity and zero density in a Hele-Shaw cell with gap $b$. 
The cell can be put in rotation with angular velocity $\Omega$ and fluid can 
be injected or sucked out of the cell through a hole at the center of 
rotation, 
with areal rate $Q$. The cases $Q > 0$ and $Q < 0$ correspond respectively to 
injecting 
or sucking fluid. As in the traditional Hele-Shaw problem, 
the flow in the viscous fluid is potential, 
${\bf v} = {\bf \nabla} \phi$, but now with a velocity potential given by 
\cite{Carrillo96}
\be
\label{phi}
\phi = - \frac{b^2}{12 \mu} \left(p - \frac{1}{2} \rho \Omega^2 r^2 \right).
\ee
Incompressibility then 
yields Laplace equation $\nabla^2 \phi = 0$ for the field $\phi$ (but 
not for the pressure).
The two boundary conditions at the interface which complete the definition 
of the moving boundary problem are the usual ones, namely, the pressure 
on the viscous side of the interface $p = - \sigma \kappa$, 
where $\sigma$ and $\kappa$ are respectively surface tension and curvature,
and the continuity condition for the normal
velocity $v_n = {\bf n} \cdot {\bf \nabla} \phi$.
The crucial difference from
the usual case is in the boundary condition satisfied by the laplacian 
field on the interface due to the last term in 
Eq.(\ref{phi}).

This problem is well suited to conformal mapping techniques 
\cite{Bensimon86}. The basic idea is to find an evolution equation
for an analytical function $z=f(\omega,t)$, which maps 
a reference region in the complex plane $\omega$, in our case 
the unit disk $|\omega| \le 1$, into
the physical region occupied by the
fluid in the physical plane $z=x+iy$, with the physical interface 
being the image of the region boundary, $|\omega|=1$. 
We consider two types of situations, one in which the 
viscous fluid is inside the region enclosed by the interface, and one 
in which it is outside.
It can be shown\cite{Carrillo96} that the 
evolution equation for the mapping $f(\omega,t)$ in the rotating case can 
be written in a compact form as
\be
\label{evolution}
{\rm Im} \{\partial_t f^* \partial_{\phi} f\} =
\frac{Q}{2 \pi} \vartheta + \frac{1}{2} \Omega^* \partial_{\phi}
{\rm H}_{\phi}[|f|^2] + d_0 \partial_{\phi} {\rm H}_{\phi}[\kappa],
\label{main}
\ee
where $\Omega^* = b^2 \Omega^2 \rho / 12 \mu $, $d_0 = b^2 \sigma / 12 \mu $,
and where we have specified the mapping function at the unit circle 
$\omega=e^{i\phi}$.
The curvature is given by 
$\kappa = -{\rm Im}\{\partial^2_{\phi}f/(\partial_{\phi}f
|\partial_{\phi}f|) \}$ 
and the Hilbert transform $H_{\phi}$ is  
defined by
\be
{\rm H}_{\phi}[g] = \frac{1}{2 \pi} {\rm P} \displaystyle \int_0^{2 \pi} 
g(\theta) {\rm cotg} \left[\frac{1}{2}(\phi - \theta) \right] d \theta.
\ee
In Eq.(\ref{evolution}), $\vartheta = +1$ and $\vartheta = -1$ 
correspond to the cases with the 
viscous fluid respectively inside or outside the interface.

Explicit time-dependent solutions of Eq.(\ref{evolution}) are 
known only for the case $d_0=0$ and $\Omega^{*}=0$. Here we report 
explicit solutions for $\Omega^{*} \neq 0$. 
A class of solutions in this context is defined by a functional form of 
the mapping (with a finite number of parameters) which is preserved 
by the time evolution.
The problem is then reduced to a set of nonlinear ODE's for 
those parameters. All solutions we have found fit into the general 
class of the rational form 
\be
\label{general}
f(\omega,t) = \omega^{\vartheta} \frac{a_0(t) + \sum_{j=1}^N a_j(t) \omega^j}
{1 + \sum_{j=1}^N b_j(t) \omega^j}, 
\ee
although not any mapping of this form is necessarily a solution. A more 
detailed study will be presented elsewhere\cite{unpublished}.
The general structure Eq.(\ref{general}) is known to yield explicit 
solutions also in the nonrotating case 
(with different ODE's for the parameters).
However, some important classes of solutions of the usual case ($\Omega^{*}=0$)
are no longer so for $\Omega^{*} \neq 0$. This is the case for instance of 
the superposition of a finite number of logarithmic terms
\cite{Dawson98} (sometimes
referred to as simple pole solutions, when considering the
derivative of the mapping rather than the mapping itself\cite{Dai91}). These 
solutions have the interesting feature of 
being free of finite-time singularities\cite{Dawson98}.
On the contrary, the 'multiple-pole' case (for the
derivative of the mapping) turns out to include solutions for both 
$\Omega^{*} = 0$ and $\Omega^{*} \neq 0$ \cite{unpublished}.
Finally,
for the polynomial case ($b_j=0$ for all $j$'s), 
the nonrotating case 
$\Omega^{*}=0$ is known to always yield  
finite-time singularities in the form of 
cusps\cite{Sarkar84,Howison85}.
We will see in the rest of this paper 
that this scenario is modified in a nontrivial way by the presence 
of rotation. 

We focus on
the role of rotation in preventing cusp formation 
in the subclass of polynomial mappings of the form
\be
f(\omega,t) = a_0(t) \omega^{\vartheta} + a_n(t) \omega^{n + \vartheta}.
\label{mapo}
\ee
For $a_n \ll a_0$ this describes a $n$-fold sinusoidal perturbation of 
amplitude $a_n$ superimposed on a 
circular interface of radius $a_0$. It is convenient to 
introduce the dimensionless 
parameter 
$\varepsilon = (n + \vartheta)a_n/a_0$. The range of physically acceptable 
values of $a_n$ and $a_0$ is given by the condition $0<\varepsilon<1$ for all
$n$. 
We also introduce a scaled mode amplitude 
$\delta=a_0 a_n$ 
which turns out to be the relevant one to characterize the interface 
instability. To see this, 
let us first compute the standard linear growth rate. 
Inserting Eq.(\ref{mapo}) into Eq.(\ref{main}) and linearizing in
$a_n$, we get,
\be
\frac{\dot{a}_n}{a_n}
 = \vartheta n \Omega^* - (\vartheta n + 1 ) \frac{Q}{2 \pi a_0^2}
- \frac{d_0}{a_0^3} n (n^2 - 1). \label{disp}
\ee
The term $-Q/2\pi a_0^2$, independent of both $n$ and $\vartheta$,
has a purely kinematic origin, associated 
to the global expansion (or contraction) of the system. 
It can easily be shown that
this quantity would be the growth rate of a mode corresponding to a  
redistribution of area given by the undistorted flow field with 
radial velocity $v=Q/2\pi r$. 
This in turn would imply $a_n(t) a_0(t) = const$\cite{unpublished}.
Accordingly, 
the marginal modes for $\delta$ (which in the rotating case may occur 
for all $n$) will be such that 
the flow field is undistorted by the interface perturbation, although 
such perturbation may grow or decay in the original variables $a_n$. 
Analogously, growth or decay of $\delta$ will correspond 
unambiguously to the stability properties of the actual 
velocity field.
In this sense it may be justified to qualify the interface instability 
as described by $\delta$ 
as 'intrinsic', as opposed to the 'morphological' one as described by 
the amplitude $a_n$.
In this way the intrinsic growth rate takes the simpler form 
\be
\label{scaled}
\frac{\dot{\delta}}{\delta}
 = \vartheta n \left(\Omega^* - Q^* \right),
\ee
where we have defined $Q^*=Q/2\pi a_0^2$ and 
have dropped the surface tension term, since hereinafter we will 
restrict to the zero surface tension case. We introduce the relevant 
dimensionless control parameter of our problem, expressing the ratio of 
centrifugal to viscous forces, as
\be
\label{P}
P = \frac{\Omega^* 2 \pi R^2}{Q} =
\frac{\pi \rho b^2 R^2 \Omega^2}{6 \mu Q}, 
\ee
where $R$ is a characteristic radius of the interface.

Eq.(\ref{scaled}) clearly exhibits the competing effects of rotation and 
injection, although their roles are not quite symmetric. In fact, 
notice that $Q^*$, which may have both signs, 
contains a dependence on $a_0$. 
In practice this means that $Q^*$ depends effectively on time.
An immediate consequence of this is that the growth of modes is not really
exponential\cite{Carrillo96} and may even be nonmonotonic. 
The asymmetry between injection and rotation shows up also in the 
fact that the sign of $Q$ determines which of the two effects dominates 
asymptotically in time. In fact, for positive injection rate the 
typical radius of the inner fluid is growing while typical interface 
velocities are 
decreasing, so centrifugal forces 
will dominate at long times. On the contrary, for negative injection rate,
typical velocities increase while typical radii decrease,  
so injection will asymptotically 
dominate over rotation. 

In view of Eq.(\ref{scaled}), the most interesting configurations will
be those in which $Q>0$, so that injection and rotation have counteracting
effects. 
In the case $\vartheta = +1$ (viscous fluid inside), which
was experimentally studied in Ref.\cite{Carrillo96}, 
rotation is always destabilizing.
A positive 
injection rate in this case tends to stabilize the circular interface. 
However, for fixed $Q$, $Q^*$ will
decrease with time, so eventually the interface will reach a radius after 
which all modes are linearly unstable. It is thus expected that, in this case,
 the formation of cusps
can only be delayed but not avoided \cite{unpublished}. 

The most interesting case from the point of view of preventing cusp formation
is $\vartheta=-1$ and $Q>0$, the usual configuration in 
viscous fingering experiments. In this case, a small rotation rate 
will slightly affect the linear instability, but 
will eventually stabilize the growth at long times,
so it is conceivable to have a nontrivial evolution starting 
from an unstable interface but not developing finite-time singularities. 

We now study the fully nonlinear dynamics of polynomial mappings.
Inserting Eq.(\ref{mapo}) into Eq.(\ref{main}) with $d_0 = 0$ we obtain 
two ordinary differential equations describing the evolution of $a_0(t)$ 
and $a_n(t)$. These can be integrated analytically and yield 
\be
&&a_0^2(t) + \vartheta(n + \vartheta) a_n^2(t) = \frac{Q}{\pi}t + k_0, \label{ana1}\\
&&a_0^{n + \vartheta}(t) a_n^{\vartheta}(t) = k_n e^{n \Omega^*t}, \label{ana2}
\ee
where $k_0$ and $k_n$ are constants to be determined by  
initial conditions, and where $n \ge 2$ for $\vartheta=+1$ and 
$n \ge 3$ for $\vartheta=-1$.

Physically acceptable solutions require that the points in the $\omega-$plane
where $\partial_{\omega} f(\omega,t) = 0$ (noninvertible) 
should lie outside the unit disk.
The occurrence of a cusp is associated to such a point crossing the unit 
circle $|\omega|=1$ at a finite time $t_c$, that is, 
\be
\left| \frac{\vartheta a_0(t_c)}{(n + \vartheta) a_n(t_c)} \right| = 1. 
\label{condit}
\ee
If we take the initial value $a_0(0)$ as the characteristic length $R$, 
which coincides with the radius of the perturbed
circle if we are in the linear regime, and define the dimensionless time 
$\tau=\Omega^*t$, condition (\ref{condit}) reads 
\be
\alpha_n \left(\frac{2 R^2 \tau_c}{P} + k_0 \right) = e^{\beta_n \tau_c}, \label{eqa}
\ee
where 
\be 
\alpha_n = \frac{(n+\vartheta)^{\frac{n}{n+2\vartheta}}}{n+2\vartheta}
k_n^{-\frac{2}{n+2\vartheta}}, \hspace{0.5cm}
\beta_n = \frac{2n}{n+2\vartheta}.
\ee
Our aim is now at finding conditions such that an initially smooth interface 
remains smooth for an infinite time. Thus we have to impose 
that Eq.(\ref{condit}) should not have any solution 
for $\tau_c>0$. The transition between 
the regions with and without cusps will be defined by the conditions that
the two members of Eq.(\ref{condit}) and their derivatives with respect 
to time are equal, such that the two curves have a common tangent. 
These two conditions allow us to eliminate $\tau_c$, and yield
\be
\label{threshold}
x \log x - x = - \alpha_n k_0
\ee
where $x=\frac{2 \alpha_n}{P \beta_n}$, and with $R=a_0(0)$ 
in Eq.(\ref{P}). We now search for solutions of 
Eq.(\ref{threshold}).
For $\vartheta = +1$ it can be proven 
\cite{unpublished}   
that this equation has no solutions, and therefore all initial conditions 
must eventually 
develop a cusp at finite time, as expected from the linear analysis. 
On the other hand, for $\vartheta = -1$ the quantity on the rhs of 
Eq.(\ref{threshold}) takes the simple form
\be
\label{rhs}
\alpha_n k_0 = \frac{n-1}{n-2} \left( 1 - \frac{\varepsilon^2}{n-1} \right)
\varepsilon^{\frac{2}{n-2}},
\ee
and a nontrivial critical line 
$P_c(\varepsilon;n)$ can be found for each $n \ge 3$. 
This implies that, in the configuration with the viscous fluid outside, 
for any initial condition
there is always a certain rotation rate above which there is no cusp formation.
The numerical determination 
of these curves is shown in Fig.1. 

\begin{figure}[h] 
\centerline{{\psfig{figure=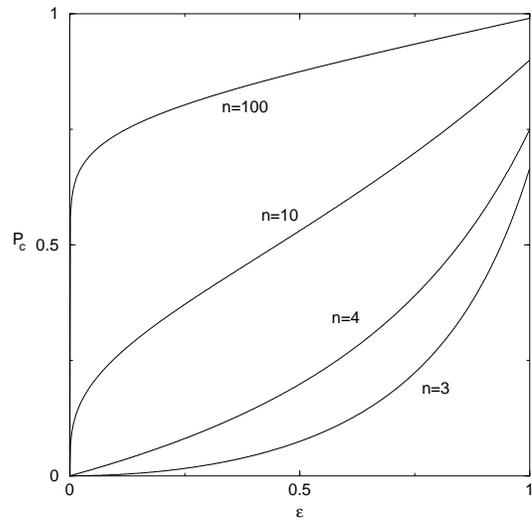,height=7cm}}}
\caption{Critical lines $P_c$ for different values of $n$. The region 
free of cusp singularities for a given $n$ is the one above the 
correponding curve.}
\end{figure}

The leading behavior for initial 
conditions in the linear regime, $\varepsilon \ll 1$, can be found by expanding 
the lhs of Eq.(\ref{threshold}) around $x=e$ and is given by
$P_c \approx \frac{n-1}{n e} \varepsilon^{\frac{2}{n-2}}$.
Notice that there are qualitative differences 
for small
values of $n$. For $n=3$ the curve starts horizontal 
at the linear level, implying that a very small rotation rate is 
sufficient to prevent cusp formation. For $n=4$ the threshold curve 
starts with a finite slope and for $n>4$ it has an infinite slope at 
$\varepsilon=0$. A more detailed description and analysis of this diagram 
will be presented elsewhere \cite{unpublished}. An example of rotation 
preventing cusp formation is shown in Fig.2.

\begin{figure}[h] 
\centerline{{\psfig{figure=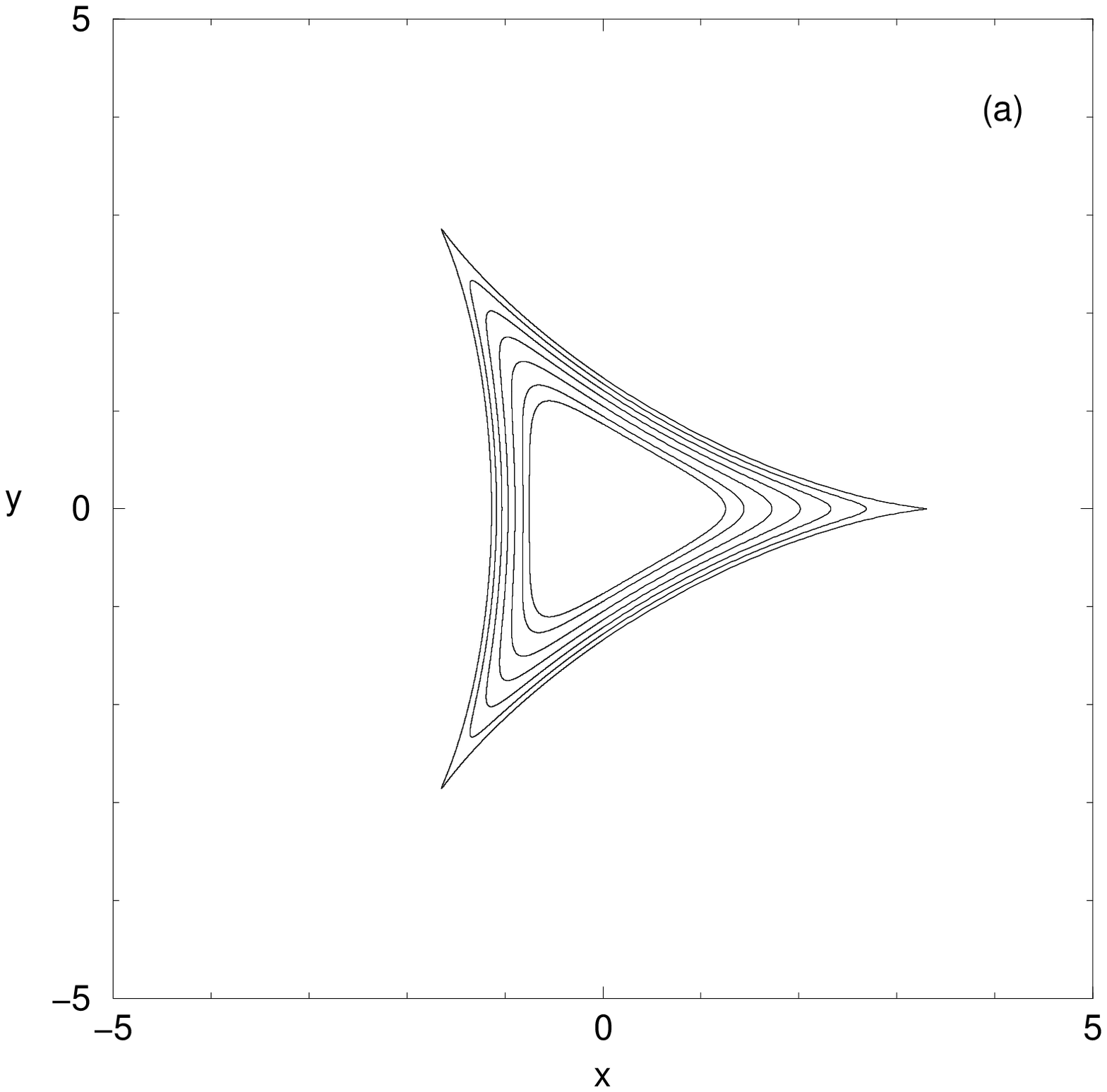,height=7cm}}}
\centerline{{\psfig{figure=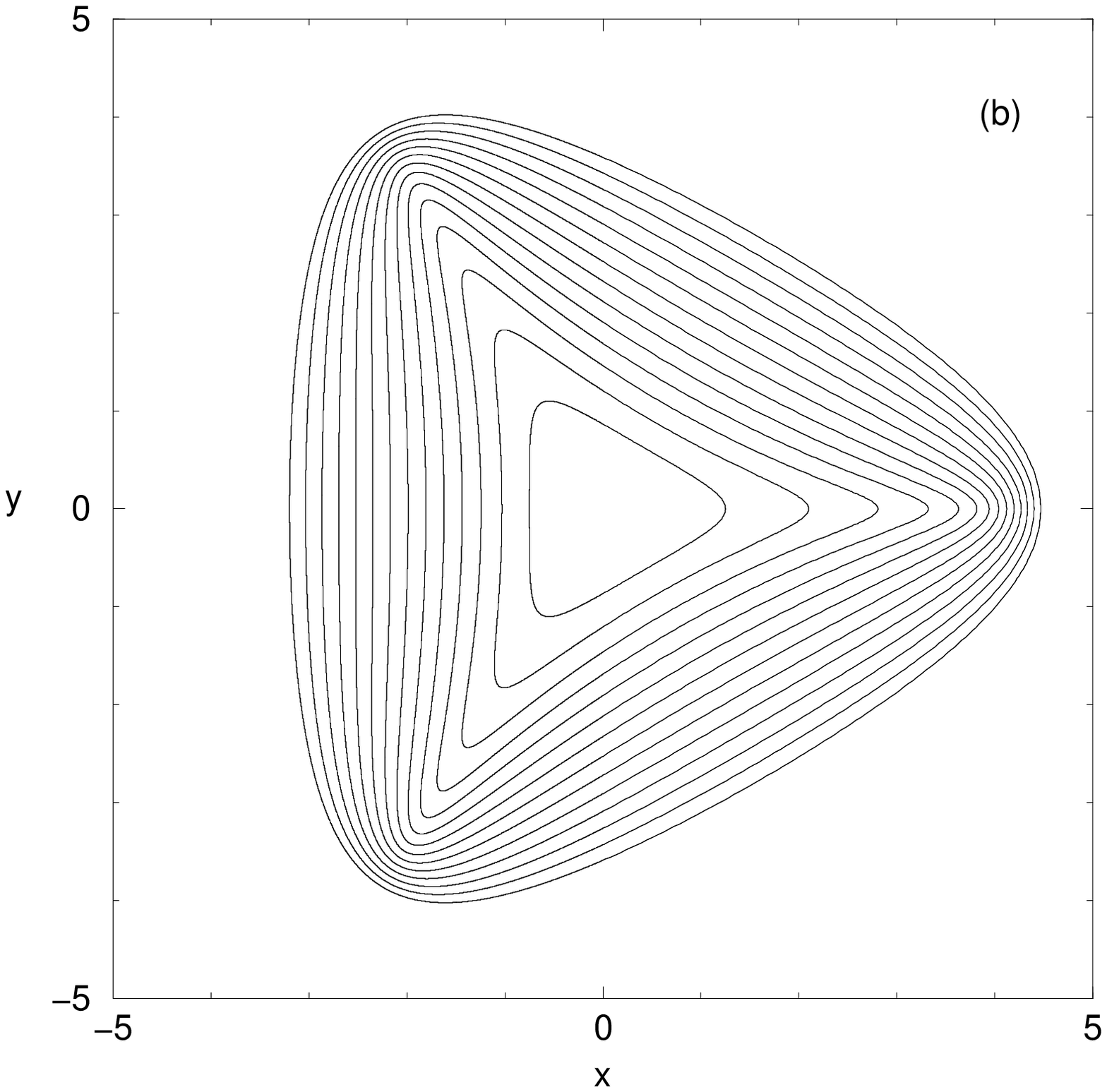,height=7cm}}}
\caption{Evolution of the interface in the case $\vartheta=-1$
(viscous fluid outside), with $n = 3$,
$a_0(0) = 1.0$, $\epsilon (0) = 0.5$, for (a) $\Omega^*= 0.025$ (cusp 
formation) and (b) $\Omega^* = 0.045$ (cusps prevented by rotation).}
\end{figure}

In Fig.1 we also see that for any given $\varepsilon$, the critical $P_c$ 
increases monotonically with $n$. If we take the limit $n \rightarrow \infty$
at fixed $\varepsilon$ we get $\alpha_n k_0 \rightarrow 1$. From 
Eq.(\ref{threshold}) this implies $x=1$ and consequently we
obtain an absolute upper bound $P_c^{max}=1$ 
for all values of $n$ and $\varepsilon$. 
This implies that,
for all initial conditions, 
there is a critical rotation rate 
\be
\label{Omegac}
\Omega_c = \left( \frac{6 \mu Q}{\pi \rho b^2 R^2} \right)^{\frac{1}{2}}   
\ee
above which cusps are always eliminated.
Although Eq.(\ref{Omegac}) has been derived for the class Eq.(\ref{mapo}),
it is expected that the existence of a certain $\Omega_c$ and the scaling 
with physical parameters given by Eq.(\ref{Omegac})
could be more general.
Notice that $P=1$ corresponds to the intrinsic marginal stability  
of the circular shape, $Q^*=\Omega^*$. Therefore, 
the sufficient condition, valid for all initial conditions of the form 
Eq.(\ref{mapo}), 
for not developing cusp singularities 
is that a circular interface with radius given by $a_0(0)$ be 
intrinsically stable, in the sense of Eq.(\ref{scaled}). 
Whether deeper consequences can be drawn in a broader context 
from this inner 
connection between the linear problem and the possibility of cusp formation 
is an interesting open question.

We acknowledge financial support by the Direcci\'on
General de Ense\~{n}anza Superior (Spain), Project PB96-1001-C02-02
and the European Commission Project ERB FMRX-CT96-0085.

\end{multicols}

\end{document}